\documentclass[conference]{IEEEtran}
\IEEEoverridecommandlockouts

\usepackage{cite}
\usepackage{amsmath,amssymb,amsfonts}
\usepackage{algorithmic}
\usepackage{graphicx}
\usepackage{textcomp}
\usepackage{xcolor}
\usepackage{microtype}
\usepackage{amsmath,amssymb,amsfonts}
\usepackage{algorithm}
\usepackage{graphicx}
\usepackage{theorem}
\usepackage{algorithmic}
\usepackage{braket}
\usepackage{physics}
\usepackage[acronym]{glossaries}
\usepackage{lineno,hyperref}
\usepackage{amsmath}
\newtheorem{definition}{Definition}

\newacronym{qkd}{QKD}{Quantum Key Distribution}
\newacronym{epr}{EPR}{Einstein–Podolsky–Rosen}
\newacronym{pki}{PKI}{Public Key Infrastructure}

\def\BibTeX{{\rm B\kern-.05em{\sc i\kern-.025em b}\kern-.08em
    T\kern-.1667em\lower.7ex\hbox{E}\kern-.125emX}}
\begin{document}

\title{Distributed Coordination Based on Quantum Entanglement

{\footnotesize (Preliminary Version)}
}

\author{\IEEEauthorblockN{Yotam Ashkenazi}
\IEEEauthorblockA{\textit{dept. of Computer Science} \\
\textit{Ben-Gurion University of the Negev}\\
Beer Sheva, Israel \\
yotamash@cs.bgu.ac.il}
\and
\IEEEauthorblockN{ Shlomi Dolev}
\IEEEauthorblockA{\textit{dept. of Computer Science} \\
\textit{Ben-Gurion University of the Negev}\\
Beer Sheva, Israel \\
dolev@cs.bgu.ac.il}
}

\maketitle

\begin{abstract}
This paper demonstrates and proves that the coordination of actions in a distributed swarm can be enhanced by using  quantum entanglement. In particular, we focus on 
\begin{itemize}
  \item Global and local simultaneous random walks, using entangled qubits that collapse into the same (or opposite) direction, either random direction or totally controlled simultaneous movements.
  \item Identifying eavesdropping from malicious eavesdroppers aimed at disturbing the simultaneous random walks by using entangled qubits that were sent at random or with predefined bases.
  \item Identifying Byzantine robots or malicious robots that are trying to gain secret information or are attacking the system using entangled qubits.
  \item The use of Pseudo Telepathy to coordinate robots' actions.
\end{itemize}
\end{abstract}


\section{Introduction}

This paper presents methods to achieve distributed coordination in a swarm of robots using quantum entanglement.
We demonstrate a new benefit of quantum mechanics (using the entanglement capabilities) in the scope of distributed secure computing. Many applications use quantum entanglement to enhance the classical algorithm capabilities. In order to achieve coordination between the robots, we use similar methods used in quantum key distribution and pseudo telepathy.

\acrfull{qkd} algorithms based on distributing entanglement photons were developed decades ago \cite{BEN84}. However, a practical experience of distributing the key to two far away participants was done a few years ago, when scientists were able to distribute entanglement photons over $1200$ kilometers using satellites \cite{SQ2017} to produce a symmetric key in two remote sites.

Quantum pseudo telepathy methods demonstrated in \cite{Brassard:1999kj}, achieved better results in several games compared to ways that do not have access to the entangled quantum system. Distributed quantum computing methods were presented recently, trying to solve and provide an overview of several interesting problems such as quantum internet and distributed quantum compiler, e.g., \cite{DBLP:journals/corr/abs-2201-03000} and  \cite{https://doi.org/10.1049/iet-qtc.2020.0002}

In addition to the above quantum techniques, we also focus on randomization, as it is a significant source in computing, particularly in distributed computing. In this paper, we employ entangled qubits to gain random coordinated actions and/or to break the symmetry. 

Randomized algorithms are used to compute a task that might have a better performance or efficiency than a deterministic non-randomized algorithm \cite{DBLP:conf/icra/AroraS17} and \cite{Yao1979SomeCQ}. Randomized algorithms might use a random sequence of bits where each bit value in the sequence is chosen randomly or pseudo-randomly. In the scope of distributed computing, randomized algorithms overcome impossible results, such as \cite{153253}, coping with situations where symmetry can not be otherwise broken. 

\section{Quantum and Classical Basic Concepts and Definitions} 
\label{def:Quantum and Classical Basic Concepts and Definitions} 

In this section, we present high-level quantum basic concepts and definitions for readers unfamiliar with quantum computing and its possible use in cryptography. More details can be found in \cite{nielsen2000quantum}. Other readers may skip this section and go directly to Section \ref{def:Preliminaries}.
\begin{itemize}
  \item Entangled qubits are two (or more) qubits with mutual influence on their values, such that one cannot describe each of them independently. For example, Bell state is defined as ${\frac{1}{\sqrt{2}}} (\ket{00} + \ket{11})$ and cannot be written as a tensor product of the two qubits. 
  \item Quantum bases are different bases that we can measure the current qubit states. The most common base is the $z$-base, which is represented by two orthogonal states $\ket{0}$ and $\ket{1}$, also called the normal basis. However, we can measure the qubits in many different bases. In this paper, the $z$-basis and the $x$-basis are used, where the $x$-basis is represented by the two states $\ket{+} =$ ${\frac{1}{\sqrt{2}}} (\ket{0} + \ket{1})$ and $\ket{-} = $ ${\frac{1}{\sqrt{2}}} (\ket{0} - \ket{1})$
  \item Quantum pseudo-telepathy 
  represents an extra capability that a pre-shared quantum entanglement qubits between several participants, might improve the strategies for several games compering to the classical strategies. 
  \item Key distribution is a family of methods to share a mutual secret key in cryptography. \acrshort{qkd} is a method to share a mutual secret key based on quantum mechanics and computation. In this paper, we use methods similar to \cite{BEN84}.
\end{itemize}

\section{Preliminaries}
\label{def:Preliminaries}
Our system settings are similar to \cite{DBLP:conf/ic-nc/AshkenaziDKOW19} with minor changes, especially in the definitions of robots moving the same tile. We abstract a region by regarding it as a board (might be an infinite board) over which the robots move. A \emph{board} is defined as a graph $G=(V,E)$, where $V$ is a set of tiles and $E\subseteq V \times V$ is a set of links. A \emph{tile} is defined as a position on the board by coordinate $(x,y)$ based on a global Cartesian coordinate system and modeled as a point in a two-dimensional Euclidean space. Tiles $t$ and $t'$ are \emph{neighboring} iff $\{t,t'\} \in E$ holds. At most one robot can occupy a tile of the board at any given instance. If two or more robots move to the same tile, the robots crash and cannot move anymore.

\section{Simultaneous Random Walks} 
\label{def:Simultaneous Random Walks}

\begin{definition}
\label{def:CoordinatedRandomWalk}
{\bf Simultaneous random walk.}
A path $P_r$ is defined as a sequence of positions of a robot $r$. $P_r=p_{r_1},p_{r_2},\ldots$, if for every $i \geq 1$, $p_{r_(i+1)}$ is reached from $p_{r_i}$ by a step of the robot $r$.
A path of random steps defines a random walk. Each participant has a random path that is not affected by other participants. A simultaneous random walk occurs when every robot's random walks are coordinated. For every $P_r$ and every $i \geq 1$, if $p_{r_(i+1)}$ is reached from $p_{r_i}$ by a step up, down, left, or right. All other $P_r` \ne P_r$ move up, down, left, or right at the same step $i$ respectively.

\end{definition}

{\bf Coordinate a random walk problem.}
\noindent
Develop an algorithm where the participants can coordinate a random walk. If robot $r_1$ moves up meaning $P_1=(i,j),(i,j+1)$. Robot $r_2$ move up as well $P_2=(k,l),(k,l+1)$. The idea is to share a random sequence between the participants, and then the participants can move in a coordinated random fashion.

{\bf A classic (no quantum) solution.} 
\noindent
There are several ways to share a random sequence. One of them, and the obvious one, is based on physical meetings where every two participants can share a secret. The participants later use the physically exchanged (and agreed on common random sequence) when executing the algorithm.

This scenario has significant drawbacks. A robot needs to predict upfront the robots that it will need to communicate with and establish a shared key in a pre-processing stage (assuming that a public key system with a certificate authority is too expensive to implement). Another drawback is the possibility of using the knowledge of the sequence and the risk of its leakage prior to the actual use of the sequence.

Every time we would like to use the random algorithm, the participants would need a new random sequence as an input to the algorithm, which implies the need for another coordination rendezvous. Our solution would like to have a random sequence with an infinite size over time whenever there is a need. 

A standard method to receive a random share sequence is to use random noise from the environment, e.g., \cite{ISPRA09}. By using this method, an (almost) truly random sequence can be achieved from the environment. Several entities may receive and analyze a common random noise (e.g., from space). However, in this scenario, an eavesdropper/ Byzantine robot/ attacker can discover/ copy the procedure for harvesting the common noise and reveal the way the other robots are going to act.
In our solution, we can, for instance, identify when a Byzantine or an attacker is eavesdropping and act accordingly.

{\bf The quantum solution.} 
\noindent
In the sequel, we propose and detail a new method to achieve distributed coordination between a swarm of robots. This can be based on one robot producing an entangled state and sending part of the state to another robot. Another option is based on a global entity (satellite, for example) sending entanglement photons to several robots. 

\noindent
Our solution suggests three ways of using quantum capabilities in the case of two robots to obtain a stream of an infinite number of random (qu)bits, while ensuring that no entity can clone or manipulate transmitted bits on their way. 
\begin{itemize}
  
\item The first option is to use predetermined bases. Using this method, the robots (and the satellite, when used) decide on predetermined bases for each measurement and measure accordingly. This option has the same drawbacks as the classical physical meeting solution. 

\item The second method uses random bases, just as done in \acrshort{qkd}. Each robot chooses a random base for each measurement. The robots then send/ broadcast their information on randomly chosen bases over another secure channel, where attackers can listen to the communication but can not modify it.

\item The third method uses quantum telepathy, based on the Mermin–Peres magic square game \cite{PhysRevLett.65.3373}. The idea is to use the game results and employ wave interference.
\end{itemize}

When using the method of distributing entangled particles from a satellite, each robot receives a part of the entangled particle infinitely often. This can also be done by one participant sending entangled qubits to another robot, and both of them measure the states.

We consider two cases of random walks. In the first one, we would like to achieve a {\it global coordinated random walk}, where the robots are located very far from each other. In this scenario, the robots may not be able to sense a common random noise from the environment and can not observe the movements of each other. Note that it is possible that the robots were close to each other in the past but later moved apart. 

In the second scenario, we would like to achieve a {\it local coordinated random walk} to prevent a collision of two robots executing random walks $P_1$ and $P_2$.
Consider the simple procedure in which a robot performs a simple random walk algorithm. The robot chooses its next move randomly with the same probability
\begin{itemize}
    \item Moving up from $(i,j)$ to $(i,j+1)$
    \item Moving down from $(i,j)$ to $(i,j-1)$
    \item Moving right from $(i,j)$ to $(i+1,j)$
    \item Moving left from $(i,j)$ to $(i-1,j)$
\end{itemize}

\begin{figure}[htbp]
\centerline{\includegraphics[scale=0.4]{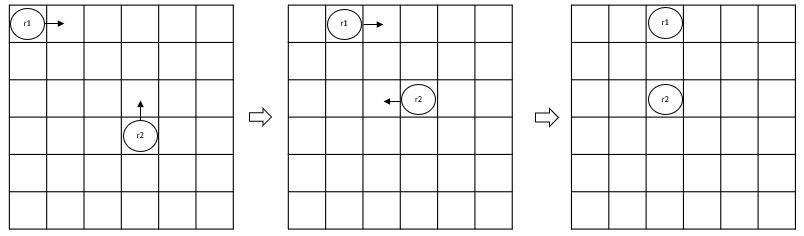}}
\caption{Robots move randomly until the distance between them is $1$}
\label{f:random}
\end{figure}

In one of the scenarios, we consider that there are two robots, $r_1$, and $r_2$, which are located very close to each other. There is a chance that $r_1$ randomly chooses to move toward $r_2$ and, at the same time, $r_2$ moves toward $r_1$. e.g., $r_1$ move right $P_1=(i,j),(i+1,j)$ and $r_2$ moves down $P_2=((i+1,j+1), (i+1,j)$ In this scenario, they may crash into each other, see Fig.~\ref{f:random}.

We can address both cases by the use of entangled qubits. The robots measure the entanglement state and act simultaneously, even if they are (possibly) very far from each other.

The robots can move in four directions. Each robot needs two qubits to decide on the next move, meaning two entangled states $\ket{\Phi1}$ and $\ket{\Phi2}$ with a total of four qubits for each step. 

The robots $r_1$ and $r_2$ measure the states, and each robot interpenetrates the measured values to a command to be executed, e.g., $\ket{00}$ up, $\ket{11}$ down, $\ket{01}$ right, and $\ket{10}$ left, where the $\ket{xy}$ represents the value measured. $r_1$ receives the first qubit of $\ket{\Phi1}$ and the first qubit of $\ket{\Phi2}$, and $r_2$ receives the second qubit of $\ket{\Phi1}$ and the second qubit of $\ket{\Phi2}$. 

\begin{figure}[htbp]
\centerline{\includegraphics[scale=0.4]{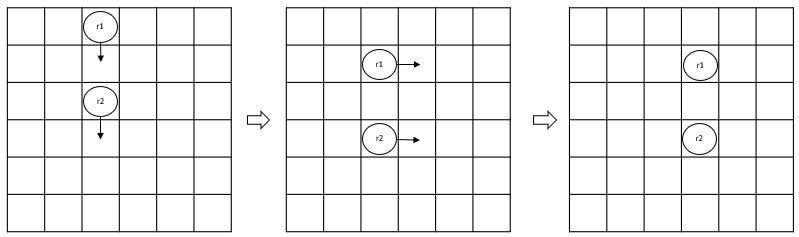}}
\caption{When the distance between the robots is $1$, the robots measure their particles from $\ket{\Phi1}$ and $\ket{\Phi2}$ and get $\ket{11}$, and they both move down. In the second step, robots measure $\ket{01}$, and they both move right.}
\label{f:entangled}
\end{figure}

We can assume that the entangled qubits are \acrfull{epr} pairs \cite{PhysicsPhysiqueFizika.1.195}, so without loss of generality, the states are both $\ket{\Phi+}$ and the robots measure on a normal basis. The robots measure their qubits and move accordingly to the result. Using this simple algorithm, assuming $r_1$ observes $\ket{01}$, $r_2$ observes the same result with a high probability and the robots move left. In case the distance between the robots is below the threshold or they want to coordinate their random walk, they can execute the algorithm above, see Fig.~\ref{f:entangled}. Therefore, they continue to move together in a random fashion and do not collide.

When using this algorithm, the robots move together forever. The rest of the paper is organized as follows. In Section \ref{def:Controlling the Robot's Movements}, we demonstrate how the centralized entity can control the robot's movements using quantum entanglement. Additionally, we consider the case where the robots can move together in a random fashion. However, a Byzantine robot or an attacker can eavesdrop on the states and predict the robots' movements. Section \ref{def:Eavesdropping Prevention section} presents a method for preventing the eavesdropping attack.
 
Previously, we used pairs of {\it EPR} entangled particles in the case of two robots. In the case of three or more robots, we can expand the $\ket{\Phi}$ state to consist of more than two qubits. e.g. for three robots, $r_1$, $r_2$, and $r_3$, we can use the state ${\frac{1}{\sqrt{2}}} (\ket{000} + \ket{111})$, and the robots need two states to create the mapping between the results and the directions. To clarify this point, let us assume that we have two states. The first state is $\ket{xyz} + \ket{xyz}$, and the second state is $\ket{abc} + \ket{abc}$. $r_1$ measures $x$ and $a$, $r_2$ measures $y$ and $b$, and $r_3$ measures $z$ and $c$. Using this method, the number of robots can be increased depending on the source of the entangled qubits, e.g., satellite, to expand the state.

\section{Controlling the Robot's Movements}
\label{def:Controlling the Robot's Movements}
The previous sections considered the case in which robots move together and execute simultaneous random walks. This section presents how a centralized entity can control the robots' movements. 

{\bf The robot swarm control problem.}
\noindent
In some cases, we would like a centralized entity (a satellite, for example) to control the robots' movements in a deterministic fashion. Say one wants to direct a swarm of drones in a specific direction. 

{\bf A classic (no quantum) solution.} 
\noindent
Controlling the robots' movements can be achieved using the same classical algorithm as in the previous section. Instead of sending a random sequence, the satellite can send specific bits which map the exact path of the robots.

{\bf The quantum solution.} 
\noindent
Controlling the robots' movements can be achieved using the same simultaneous random walks quantum algorithm. However, instead of sending a random {\it EPR} state, the satellite can send an entangled state in the form of $\ket{00}$ or $\ket{11}$. In this case, using the same conditions as above, the centralized entity can decide on the complete path of the robots. The robots continue to execute the algorithm and can not identify their movements as being predefined by the centralized entity. In addition, the centralized entity can control the robot's movement by using a different state, so each robot moves in a different direction. The centralized entity can prevent the situation in which robots stay closed forever while the robots do not move in a random fashion.

\section{Avoid Robots Colliding in a Random Fashion}
\label{def:Avoid Colliding in a Random Fashion}

{\bf The collision avoidance problem.}
\noindent
The previous section considered the case to avoid collision in a deterministic way. In this section, the robots avoid colliding and still move in a random fashion.

{\bf A classic (no quantum) solution.} 
\noindent
It is not trivial to solve the problem using a classical algorithm. The centralized entity can use one of the methods to share random sequence as demonstrated in Section \ref{def:Simultaneous Random Walks}. However, if the centralized entity sends the same sequence to the robots, the robots keep moving together forever. One solution for the problem is when the centralized entity sends a different sequence to each of the robots. The centralized entity measures the two random bits from the sequence of $r_1$, calculates all other options for two bits to $r_2$ and chooses one option randomly. $r_1$ and $r_2$ receive the bits and act accordingly. 

{\bf The quantum solution.} 
\noindent
The centralized entity creates a random state with fewer options, so the robots continue to move in a random fashion without the probability of colliding. This can be done by sending two different {\it EPR} states where the robots move in a random direction but not toward each other. For example, if two robots are located at a distance one from each other, then the centralized entity can send the first pair ${\frac{1}{\sqrt{2}}} (\ket{00} + \ket{11})$ and ${\frac{1}{\sqrt{2}}} (\ket{01} + \ket{10})$ as the second pair. In this case, the options for robot $r_1$ and robot $r_2$ are:
\begin{itemize}
  \item $r_1$ and $r_2$ measure the left qubit from the first pair $\ket{\Phi1}$ and the left qubit from the second pair $\ket{\Phi2}$, such that $r_1$ observes ${00}$ and  $r_2$ observes ${01}$. $r_1$ moves up, and $r_2$ moves right.
  \item $r_1$ and $r_2$ measure the left qubit from the first pair $\ket{\Phi1}$ and the right qubit from the second pair $\ket{\Phi2}$, such that $r_1$ observes ${01}$ and  $r_2$ observes ${00}$. $r_1$ moves right, and $r_2$ moves up.
  \item $r_1$ and $r_2$ measure the right qubit from the first pair $\ket{\Phi1}$ and the left qubit from the second pair $\ket{\Phi2}$, such that $r_1$ observes ${10}$ and  $r_2$ observes ${11}$. $r_1$ moves left, and $r_2$ moves down.
  \item $r_1$ and $r_2$ measure the right qubit from the first pair $\ket{\Phi1}$ and the right qubit from the second pair $\ket{\Phi2}$, such that $r_1$ observes ${11}$ and  $r_2$ observes ${10}$. $r_1$ moves down, and $r_2$ moves left.
\end{itemize}
Another option is that the centralized entity can send the pairs ${\frac{1}{\sqrt{2}}} (\ket{01} + \ket{10})$ and ${\frac{1}{\sqrt{2}}} (\ket{00} + \ket{11})$. In this case, the options for robot $r_1$ and robot $r_2$ are:
\begin{itemize}
  \item $r_1$ and $r_2$ measure the left qubit from the first pair $\ket{\Phi1}$ and the left qubit from the second pair $\ket{\Phi2}$, such that $r_1$ observes ${00}$ and  $r_2$ observes ${10}$. $r_1$ moves up, and $r_2$ moves left.
  \item $r_1$ and $r_2$ measure the left qubit from the first pair $\ket{\Phi1}$ and the right qubit from the second pair $\ket{\Phi2}$, such that $r_1$ observes ${01}$ and  $r_2$ observes ${11}$. $r_1$ moves right, and $r_2$ moves down.
  \item $r_1$ and $r_2$ measure the right qubit from the first pair $\ket{\Phi1}$ and the left qubit from the second pair $\ket{\Phi2}$, such that $r_1$ observes ${10}$ and  $r_2$ observes ${00}$. $r_1$ moves left, and $r_2$ moves up.
  \item $r_1$ and $r_2$ measure the right qubit from the first pair $\ket{\Phi1}$ and the right qubit from the second pair $\ket{\Phi2}$, such that $r_1$ observes ${11}$ and  $r_2$ observes ${01}$. $r_1$ moves down, and $r_2$ moves right.
\end{itemize}

\begin{figure}[htbp]
\centerline{\includegraphics[scale=0.33]{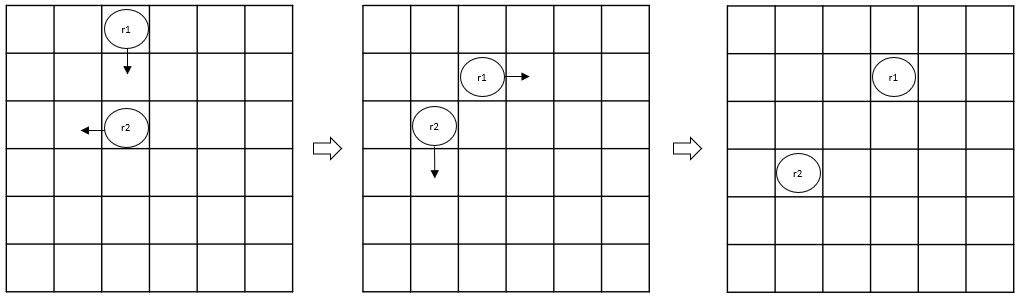}}
\caption{In the first step, robots measure the fourth option from case $1$ ($r_1$ observes ${11}$ and moves down. $r_2$ observes ${10}$ and moves left). In the second step, the robot measures the second option from the second case ($r_1$ observes ${01}$ and moves right, and $r_2$ observes ${11}$ and moves down).}
\label{f:increase_distance}
\end{figure}

In the cases above, the distance between the robots can increase or remain identical with a positive probability, see Fig.~\ref{f:increase_distance}

\section{Eavesdropping Prevention}
\label{def:Eavesdropping Prevention section}

{\bf The eavesdropping prevention problem.}
\noindent
In the previous sections, we presented a method of distributed coordination. An eavesdropper can easily attack this method by measuring the random sequence before/together with the robots. In our case, we have four identities; the centralized entity $c$ sending the random sequence, $r_1$ and $r_2$ receiving the sequence, and an attacker $eve$ trying to gain information on the random sequence or the next robot's movements.

{\bf A classic (no quantum) solution.} 
\noindent
In case the entities share a secret or have a \acrfull{pki}, the obvious and most straightforward method to avoid eavesdropping is to use encryption. Consider the case where all the information is encrypted and $eve$ does not have the secret, no eavesdropping can be done. 

{\bf The quantum solution.} 
\noindent
We extend the quantum algorithm to be resilient to eavesdropping attacks by sending the quantum states in one of several bases. We obtain very high security using our solutions, the same as the secure method for quantum key distribution, e.g., \cite{Nature_cryptography}.

The first and easy option is to use predefined bases. This solution has the same drawbacks as the physically meeting solution, and we decided to present it despite it. The participants can use predefined bases, so each of the $c$, $r_1$, and $r_2$ have the same sequence of bases and, each state can be measured on the $z$ basis or $x$ basis. 

$c$ creates the state $\frac{1}{\sqrt{2}}(\ket{00} + \ket{11})$ for the $x$ basis and $\frac{1}{\sqrt{2}} (\ket{++} + \ket{--})$ for the $z$ basis and sends the entangled states to $r_1$ and $r_2$. $r_1$ and $r_2$ measure (separately) their qubits on the predefined basis. In this case, $c$, $r_1$, and $r_2$ measure the state on the same basis, and this is a valid measurement. 

Another case is to use randomized bases. $c$ chooses a random base, $z$ basis or $x$ basis and creates the state $\frac{1}{\sqrt{2}}(\ket{00} + \ket{11})$ or $\frac{1}{\sqrt{2}} (\ket{++} + \ket{--})$ respectively. For each received qubit, each robot $r_1$ and $r_2$ choose a random basis, $z$ basis or $x$ basis and measures the qubit state using the selected basis. After several measurements, $c$, $r_1$, and $r_2$ publish their selected bases in an authenticated secure channel. A valid measurement is when $c$, $r_1$, and $r_2$ choose the same basis for each measurement. If the basis is selected in a random fashion, the probability of the same basis is $\frac{1}{4}$. 

When $eve$ is not active, $c$, $r_1$ and $r_2$ keep only the valid measurements, and $c$, $r_1$, and $r_2$ have the same values. 
At this point, the parties $c$, $r_1$, and $r_2$ use an authenticated secure channel where $eve$ has access to the data in the channel but can not change it. This algorithm still required a shared secret key or \acrshort{pki} same as the classical algorithm. However, it does not require encryption to create the authenticated secure channel. 

In the case of $eve$ being active, we would like to prevent $eve$ from eavesdropping on the states. Without loss of generality, for all the valid measurements, consider the case where $eve$ measures the state before $r_1$ (or $r_2$) and returns the state after the measurement to $r_1$. If $eve$ measures the state with the same basis as $r_1$, $eve$ and $r_1$ (and $c$, and $r_2$) measure an identical value. If $eve$ measures the state with another basis and then sends the state to $r_1$, $r_1$ might measure a different result from $r_2$. In order to identify $eve$, $r_1$ and $r_2$ can publish several valid measurement results. If the measurement results are not identical (more than an error rate), $c$, $r_1$, and $r_2$ can assume $eve$ eavesdropped on several states, and the measurements are invalid. Using this method, honest participants can identify if an eavesdropping attack was executed with a high probability. 

\section{Identify Byzantine Robots Using Entangled Qubits}
{\bf Identify Byzantine Robots problem.}
\noindent
In this section, we present a method of identifying Byzantine robots based on entangled qubits, where the non-Byzantine robots agreed on predefined bases. This is an easier problem than the previous section, although it uses a different solution from the known solution presented in the \acrshort{qkd}.

{\bf The quantum solution.} 
\noindent
Consider the case of simultaneous random walks, where the centralized entity and all the $r_i$ non-Byzantine robots agreed on predefined bases. This scenario can be done using a physical meeting of all the participants. However, it can be done by physical meetings of two identities at a time, meaning the centralized entity can meet $r_1$ and agree on the bases. Then, $r_1$ can meet $r_2$ and transfer the information about the bases until all $r_i$ have the same bases. In this scenario, the centralized entity does not know which of the robots have predefined bases and which of the robots do not have them. The robots that do not have the bases consider Byzantine robots $b_j$ and have no knowledge of the bases.

For each step, the centralized entity creates two entangle states with $i+j$ qubits each, so each robot (Byzantine and non-Byzantine) receives two qubits (in order to move). During every step, all the non-Byzantine robots measure their qubits and act accordingly. The Non-Byzantine robots move in the same direction, as they all measure on the same base and receive the same results. 

The Byzantine robots have several methods to decide on their next move. The first method is to guess the base and measure the qubits. The chance to move to the correct location using the first method is $50\%$, as the predefined bases were chosen in a random fashion from two options. Another method is to decide on a random direction and move accordingly. When using the second method, each Byzantine robot has a $25\%$ chance to move with the honest robots. Using the first two methods, the probability that a Byzantine robot guesses all the correct movements for a long time is negligible.
The third method is to wait until the non-Byzantine robots start to move and follow them. In the third method, it is easy to identify the Byzantine. The non-Byzantine robots can synchronize the time of their movements and, by that, can determine which of the robots delay and recognize them as Byzantine.

\section{Coordinated Random Walk with More Than Two Robots}
{\bf The problem.}
\noindent
In Section \ref{def:Simultaneous Random Walks}, we presented an algorithm to achieve coordination between the robots. This section presents a method to achieve coordination in a multi (more than two) robot swarm. 

{\bf A classic (no quantum) solution.} 
\noindent
In case all the robots receive the same random sequence, the same algorithm presented in Section \ref{def:Simultaneous Random Walks} can be executed here.

{\bf The quantum solution.} 
\noindent
If we use random bases with multi (more than two) robots, the solution is more sophisticated. The obvious and trivial methods can work, but if the number of robots increases, the probability that all the robots measure the same state decrease dramatically. e.g., the probability that all $c$, $r_1$, \ldots $r_n$ choose the same basis from the two options, $z$ basis or $x$ basis, is ($\frac{1}{2}$)$^n$. This method is inefficient and can cause many ``invalid`` measurements. 

\begin{figure}[htbp]
\centerline{\includegraphics[scale=0.4]{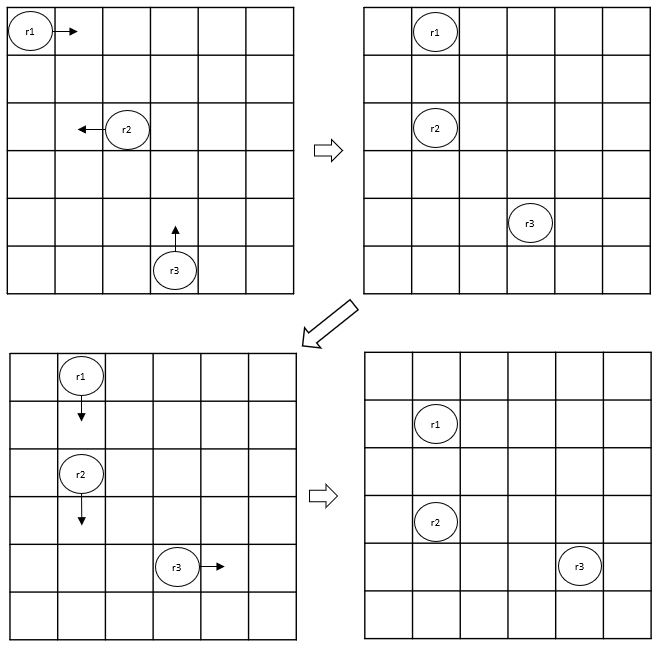}}
\caption{$r_1$, $r_2$, and $r_3$ move randomly. When the distance between robots $r_1$ and $r_2$ is $1$, they use the shared valid measurement and move down. $r_3$ continues to move randomly.}
\label{f:3_robots}
\end{figure}

In order to improve the method above, each robot can execute the same algorithm as in Section \ref{def:Eavesdropping Prevention section}. However, instead of ignoring all the measurements where the basis is not the same for all the robots, each robot stores the results where the measurement is equal between a subset of the robots and $c$. e.g., if we have three robots $r_1$, $r_2$ and $r_3$. Assume $c$, $r_1$ and $r_2$ measure on the same basis, while $r_3$ measures on a different basis. $r_1$ stores the result of this measurement for only an equal result with $r_2$ (and $r_2$ stores the result of this measurement for only $r_1$). In case some operations involve $r_1$ and $r_2$ only, they can still use the measurement results, see Fig.~\ref{f:3_robots} 

\section{Quantum Pseudo Telepathy Among Swarming Robots}
\label{def:magic square game}

{\bf The problem.}
\noindent
In this section, we present a method to use quantum telepathy to achieve coordination between the robots using the same idea of Mermin–Peres magic square game described in \cite{PhysRevLett.65.3373} and \cite{PERES1990107}. The game includes a $3 \times 3$ board and each tile consists of $1$ or $-1$. The first player returns the line values where the multiple of each tile in the row is $1$. 
The second player returns the values of a column where the multiple of the tiles in the column is $-1$. 
The players can share information before the game begins but can not share any information later. 

The centralized entity sends a line number to the first player and a column number to the second player. The players win if the number in the tile shared by their row and column is the same.

{\bf The quantum solution.} 
\noindent
It is easy to prove that if the players do not know the line and column numbers in advance, the best win probability without using a quantum entanglement is $8/9$. However, if the two players share two quantum entanglement states, they can win with a probability of $1$.

The robots can decide on predefined bases for each of the tiles on the board and measure the states using the relevant bases. In our scenario, the two robots can achieve coordination in case the centralized entity publishes the information about which row and column numbers were selected. The two robots have the same result in the shared tile. 

Another option to consider is if the robots send their results (one robot sends the row result and the second sends the column result) to a board with $9$ sensors arranged in a $3 \times 3$ structure. The sensors receive the results, and if a wave interference occurs, the sensor executes an action. As the results in the shared tile are equal, only one relevant sensor (the sensor in the chosen line and row) identifies the wave interference.

Although this algorithm needs two quantum entanglement states, which is less efficient than the previous method, the players have additional information they can use later in this method. In addition, the first robot knows that the multiple of each tile in the chosen row is $1$, and the second robot knows that the multiple of each tile in the chosen row is $-1$. 

\section{Conclusions}
We demonstrated the usage and benefits of using quantum entanglement to achieve simultaneous random walks between robots. In addition, we presented several methods to identify Byzantine robots to eavesdrop and disturb the execution of the random walk using quantum phenomenons and described ways to extend our algorithm to a multi-robots environment. Interestingly, while designing our algorithms, new multiple participant's \acrshort{qkd} techniques were established.

\bibliographystyle{unsrt}
\bibliography{refs}

\end{document}